\documentclass[11pt]{article}
\usepackage[cp1251]{inputenc}
\usepackage[english]{babel}   
\usepackage{amsfonts}
\usepackage{graphics}
\usepackage{amssymb}
\usepackage{amsmath}
\usepackage{epsfig}
\usepackage{cite}
\begin{document}
\sloppy

\begin{center}
\textbf{''Optical''  Spin Rotation  Phenomenon and Spin Filtering
of Antiproton (Proton, Deuteron) Beams in a Nuclear Pseudomagnetic
Field of a Polarized Nuclear Target: the Possibility of Measuring
the Real Part of the Coherent Zero-angle Scattering  Amplitude}
\end{center}

\begin{center}
\textbf{V.G. Baryshevsky}
\end{center}

\begin{center}
\textit{Research Institute for Nuclear Problems,
\\11 Bobruiskaya str., 220030, Minsk, Belarus,\\ e-mail: bar$@$inp.minsk.by, v$\_$baryshevsky$@$yahoo.com}
\end{center}

\textbf{Introduction}

The phenomena of interference, diffraction and refraction of light
are well known even to lycee and college students. A great variety
of their applications are described in school and university
manuals and popular science books \cite{vfel_Born,B2,B3,B4}.
Centuries-long argument about the nature of light:  whether light
is a wave or a particle, finally led to  creation of quantum
mechanics and extension of the wave conception to the behavior of
any particles of matter. As a result, optical concepts and notions
were also introduced for describing interaction of particles with
matter, nuclei and one another \cite{2,rins_1,th_11}. In
particular, widely used nowadays is diffraction of electrons and
neutrons  by crystals, which are, in fact, natural diffraction
gratings. Neutron interferometers were designed
\cite{rins_79}. It was found out that scattering of particles
by nuclei (and by one another) is in many cases similar to
scattering of light by a drop of water (the optical model of the
nucleus).

A study of interaction between light and matter showed that besides
frequency and propagation direction, light waves are characterized by polarization.

The first experiment, which observed a phenomenon caused by the polarization of light,
was carried out in 1669 by E. Bartholin,  who discovered the double refraction of a light ray
by Iceland spar (calcite). Today it is common knowledge that in the birefringence effect, the
stationary states of light in a medium are the states with linear polarization parallel
or perpendicular to the optical axis of a crystal. These states have different indices of refraction
and move at different velocities in a crystal. As a result in crystals, for example, circularly
polarized light  turns into linearly polarized and vice versa \cite{vfel_Born}.

Another series of experiments was performed by D.F. Arago in 1811 and J.B. Biot in 1812. They
discovered the phenomenon of optical activity, in which the light polarization plane rotates with
the light passing through a medium. In 1817 A. Fresnel established that in an optically active medium
rotating the polarization plane, the stationary states are the waves with right-hand and left-hand
circular polarizations, which, as he found out in 1823, move in a medium at different velocities
(i.e., propagate with different indices of refraction ), thus causing polarization plane rotation. Let us also
recall the effect of light polarization plane rotation in matter placed to a magnetic field, which
was discovered by Faraday, and the birefringence effect in matter placed to an electric field
(the Kerr effect).

The above-mentioned phenomena and various other effects caused by
the presence of polarization of light and optical anisotropy of
matter have become the subjects of intensive studies and found wide applications.
In the final analysis, the microscopic mechanism behind the
optical anisotropy of matter is due to the
dependence of the electromagnetic wave scattering
by an atom (or molecule) on the wave polarization (i.e., on the photon spin),
and to the electron bindings in
atoms and molecules. Beyond the optical spectrum, when the photon
frequency appears to be much greater than characteristic atomic
frequencies, these bindings become negligible, and the electrons can
be treated as free electrons. As a result, the effects caused by
optical anisotropy of matter, which are studied in optics, rapidly
diminish, becoming practically unobservable when the wavelengths
are smaller than $10^{-8}$ cm.

Moreover, there is a widespread belief that it is only possible to speak of the refraction of
light and to use the concept of the refraction index of light
in matter because the wavelength of light
($\lambda \approx 10^{-4}$ cm) is much greater than the distance
between the atoms of matter $R_a$ ($R_a\approx 10^{-8}$ cm), since
only in this case ($\lambda \gg R_a$) matter may be treated as a
certain continuous medium. As a consequence, in a short-wave
spectral region where the photon wavelength is much smaller than
the distance between the atoms of matter, the effects similar to
the Faraday effects and birefringence, which are due to refraction,
should not occur. However, such a conclusion  turned out to be incorrect. The
existence of the refraction phenomena does not appear to be
associated with the relation between the wavelength $\lambda$ and the
distance between atoms (or scatterers). Even at high photon
energies, when the wavelength is much smaller than $R_a$, the
effects due to wave refraction in matter can be quite appreciable.
Thus, for example, when a beam of linearly polarized
$\gamma$-quanta with the energies greater than tens of kiloelectronvolts
(wavelengths smaller than $10^{-9}$ cm) passes through matter with
polarized electrons, there appears rotation of the polarization plane of
$\gamma$-quanta, which is  kinematically analogous to the Faraday effect
(\cite{51,52,54,55,56}).  Moreover, with growing  energy of the $\gamma$-quantum
(decreasing wavelength of the $\gamma$-quantum ) the effect increases, attaining its maximum
in the megaelectronvolt energy range.
Unlike the Faraday effect, which is due to
the electron bindings in atoms, for
$\gamma$-quanta, electrons  may be treated as free electrons. The effect of
polarization plane rotation in this case is due to the
quantum-electrodynamic radiative corrections to the process of
scattering of $\gamma$-quanta by an electron, which are lacking in classical
electrodynamics.

In a similar manner, the propagation in matter of the de Broglie
waves, which describe the motion of massive particles, may be
characterized by the index of refraction \cite{rins_1,1}. In this
case, the index of refraction also characterizes particle motion
in matter, even at high energies, for which the de Broglie
wavelength $\hbar/m v$ ($m$ is the particle mass, $v$ is its
velocity, in the case of relativistic velocities $m$ stands for
the relativistic mass $m\gamma$, $\gamma$ is the Lorentz factor)
is small in comparison with the distance between the atoms
(scatterers). Furthermore, it turns out that for particles with
nonzero spin, there exist the phenomena analogous to light
polarization plane rotation and birefringence
\cite{24,110,232,journ_COSYexperiment}.  In this case such
phenomena of quasi-optical activity of matter ("optical"
anisotropy of matter) are due not only to electromagnetic but also
to strong and weak interactions.

The investigations in this field were initiated in 1964 with the
publication of two papers: by F. Curtis Michel (1964) and by V.
Baryshevsky and M.Podgoretsky (1964).  F. Curtis Michel (1964)
\cite{87} predicted the existence of spin "optical rotation" due
to parity nonconserving weak interactions (the phenomenon
was experimentally revealed \cite{92}  
 and is used for studying parity nonconserving weak
interactions between neutrons and nuclei). 
 V. Baryshevsky and M.Podgoretsky (1964) \cite{24} predicted the existence of the phenomenon of quasi-optical spin
rotation of the neutron moving in matter with polarized nuclei, which is caused by strong interactions,
and introduced the concept of a nuclear pseudomagnetic field (neutron spin precession
in a pseudomagnetic field of matter with polarized nuclei). The concept of a nuclear
pseudomagnetic field and the phenomenon of neutron spin precession in matter with
polarized nuclei were experimentally verified by Abragam's group in France (1972)
\cite{25} and Forte in Italy (1973) \cite{26}.

Further analysis showed that the effects due to optical
activity of matter, which we consider in optics, are, in fact, the
particular case of coherent phenomena emerging when polarized
particles pass through matter with nonpolarized and polarized
electrons and nuclei \cite{164,Nuclear_optics}.
It was found out, in particular, that at high energies of particles
(tens, hundreds and thousands of gigaelectronvolts), the effects of
"optical anisotropy" are quite significant, and they may become
the basis of unique methods for the investigation of the structure
of elementary particles and their interactions.

In \cite{109,110,232,nim06_PR+,21a,journ_8rot,EDM}, it was shown,
in particular, that by measuring the spin rotation angle of
high-energy particles, which pass through a polarized
(nonpolarized) target (transmission experiment), one can measure
the spin-dependent real part of the elastic zero-angle scattering
amplitude.

Currently, there is a lot of discussion about the possibility to
produce polarized beams of antiprotons during their passage
through a polarized gas target placed in a storage ring
\cite{24a,PAX,PAX1,mil,mil1,nik,nik1,nik2,R}.

In the present paper we would like to highlight the possibility to
measure in such experiments not only spin-dependent total
cross-sections of antiproton scattering by the proton (deuteron),
but also the spin-dependent real part of the coherent zero-angle
scattering amplitude in the process of production of a polarized
beam of antiprotons. This means that even when an unpolarized beam
of antiprotons passes through a polarized hydrogen (deuteron)
target, one can measure the spin-dependent real part of the
coherent zero-angle scattering amplitude.


\section{"Optical" spin precession of relativistic particles in polarized targets}

Let us remember the consideration of the effects of "optical" spin
rotation arising when high-energy particles pass through matter
with polarized nuclei
\cite{109,110,nim06_PR+,21a,journ_8rot,EDM,R1}.

To be more specific, we shall consider refraction of relativistic
protons (antiprotons)  in matter. To begin with, let us analyze
scattering by a particular center. The asymptotic expression for a
wave function describing scattering of relativistic particles in
the field of a fixed scatterer far from the scatterer can be
represented in the form \cite{2,1}
\begin{equation}
\label{18.1}
\Psi=U_{E,\vec{k}}e^{ikz}+U^{\prime}_{E^{\prime}\vec{k}^{\prime}}\frac{e^{ik^{\prime}r}}{r},
\end{equation}
where $U_{E\vec{k}}$ is the bispinor amplitude of the incident
plane wave; $U^{\prime}_{E^{\prime}\vec{k}^{\prime}}$ is the
bispinor describing the amplitude of the scattered wave; $E$ and
$\vec{k}(E^{\prime},\vec{k}^{\prime})$ are the energy and the wave
vector of the incident (scattered) wave.

According to \cite{23}, the bispinor amplitude is fully determined
by specifying a two-component quantity --- a three-dimensional
spinor  $W$, which is a non-relativistic wave function in the
particle rest frame. For this reason, the scattering amplitude,
i.e., the amplitude of a divergent spherical wave, in (\ref{18.1})
similarly to the non-relativistic case, can be defined as
a two-dimensional matrix of $\hat{f}$ by the relation
$W^{\prime}=\hat{f}W$, where $W^{\prime}$ is the spinor
determining the bispinor
$U^{\prime}_{E^{\prime}\vec{k}^{\prime}}$. Thus determined
scattering operator is quite similar to the operator scattering
amplitude in the non-relativistic scattering theory allowing for
spin.

As a result, deriving the expression for the  index of refraction by
analogy with a non-relativistic case (see \cite{Nuclear_optics,109,nim06_PR+},
we obtain the following expressions for the wave function of a
relativistic neutron (proton) in a medium
\begin{equation}
\label{18.2} \Psi= \frac{1}{\sqrt{2E}}\left(\begin{array}{cc}
\sqrt{E+m}e^{ik\hat{n}z} & W\\
\sqrt{E-m}(\vec{\sigma}\vec{n})e^{ik\hat{n}z} & W
\end{array}\right),
\end{equation}
where
\begin{equation}
\label{18.3} \hat{n}=1+\frac{2\pi\rho}{k^{2}}\hat{f}(0)
\end{equation}
is the operator refractive index, $\hat{f}(0)$ is the operator
amplitude of coherent elastic zero-angle scattering by a polarized
scatterer; $\vec{\sigma}$ is the vector made up of the Pauli
matrices; $n=\vec{k}/\vec{k}$.

Using $\vec{\sigma}$, $\vec{J}$, $\vec{n}$ ($\vec{J}$ is the
nuclear spin operator), we may write  the amplitude of coherent
elastic forward scattering by a polarized nucleus in the general
case of strong, electromagnetic and space (P)- and time
(T)-invariance violating  weak interactions
\cite{Nuclear_optics,110,journ_8rot,journ_ANKEproposal}

According to (\ref{18.2}), the spinor $W^{\prime}$ defining the
spin state of a particle in the rest frame after passing the path
length $z$ in the target has the form
\begin{equation}
\label{18.4} W^{\prime}=e^{ik\hat{n}z}W.
\end{equation}
Note that (see  \cite{109,110,nim06_PR+}),  $\hat{n}$ can be written as
\begin{equation}
\label{18.5} \hat{n}=n_{0}+\frac{2\pi\rho}{k^{2}}(\vec{\sigma}
\vec{g}),
\end{equation}
where $n_{0}$ is the $\vec{\sigma}$-independent part of $n$,
\begin{equation}
\label{ins_18.5}
n_{0}=1+\frac{2\pi\rho}{k^{2}}(A+A_{3}\vec{n}\vec{n}_{1}+B_{4}\vec{n}\langle\vec{J}\rangle+\ldots).
\end{equation}
\begin{equation}
\label{ins1_18.5}
\vec{g}=A_{1}\langle\vec{J}\rangle +A_{2}\vec{n}(\vec{n}\langle\vec{J}\rangle)+B\vec{n}+B_{1}[\langle\vec{J}\rangle\vec{n}]+\ldots,
\end{equation}
where $\langle\vec{J}\rangle=Sp\,\hat{\rho}_J\vec{J} $, $\hat{\rho}_J$ is the spin density matrix of the target. Assume that $\langle\vec{J}\rangle=J\vec{p}$, where $\vec{p}$ is the target polarization vector, $\vec{n}_1$ has the components $n_i=\langle Q_{ik}\rangle n_k$, $\langle Q_{ik}\rangle$ is the quadrupolarization tensor of the target. It is non-zero when the nuclei spin $J\ge 1$. In a polarized target with  nuclei of spin $J=\frac{1}{2}$,  it is absent.
Suppose that particle absorption can be neglected, as a consequence, $\vec{g}$ is a real
vector. With the help of (\ref{18.5}), represent (\ref{18.4}) as
follows ($\vec{j}_{g}=\vec{g}/|\vec{g}|$)
\begin{equation}
\label{18.6}
W^{\prime}=e^{ikn_{0}z}e^{i\frac{2\pi\rho}{k}(\vec{\sigma}\vec{j}_{g})|\vec{g}|z}W.
\end{equation}
Now recall that the operator of spin rotation through an angle
$\vartheta$ about a certain axis, characterized by a unit vector
$\vec{j}$ has the form
\begin{equation}
\label{18.7} \hat{T}=e^{i\frac{\vartheta}{2}\vec{\sigma}\vec{j}}.
\end{equation}
Comparing (\ref{18.4}), (\ref{18.5}) and (\ref{18.6}), we obtain
that in the case in question the operator
\[
exp\left\{i\frac{2\pi\rho}{k}|\vec{g}|(\vec{\sigma}\vec{j}_{g})z\right\}
\]
acts as a spin rotation operator of a particle in its rest
frame. The rotation angle is
\begin{equation}
\label{18.8}
\vartheta=\frac{4\pi\rho}{k}|\vec{g}|z=k(n_{\uparrow\uparrow}-n_{\downarrow\uparrow})z,
\end{equation}
where the quantization axis is chosen along $\vec{j}_{g}$.

To be more specific, we shall further analyze the effect of spin
precession in a polarized target due to strong interactions. In
view of (\ref{18.5}),(\ref{ins_18.5}), (\ref{ins1_18.5}), and  (\ref{18.8}), for the angle of
particle spin rotation about the direction $\vec{j}_{g}$ we obtain
\begin{equation}
\label{18.9}
\vartheta=\frac{2\pi\rho}{k}\texttt{Re}(f_{\uparrow\uparrow}-f_{\uparrow\downarrow})z=
\frac{4\pi\rho}{k}|\texttt{Re} A_{1}\langle\vec{J}\rangle+ \texttt{Re} A_{2}\vec{n}(\vec{n}\langle\vec{J}\rangle)|z.
\end{equation}
To answer a question about how the spin precession
angle $\vartheta$ of a relativistic particle  in a polarized target
depends on the particle energy, let us remember that at scattering by a
potential, the Dirac equation for ultrarelativistic particles
reduces to the equation similar to a non-relativistic
Schr\"odinger equation, where the particle mass $M$ stands for its
relativistic mass \cite{23}, i.e., $M=\gamma m$, where $m$ is the
particle rest mass, $\gamma$ is its Lorentz factor. As the
amplitude of  particle scattering by a potential is proportional
to the particle mass, then the amplitude for a relativistic particle may be written as
\begin{equation}
\label{18.10} f(E, 0)=\gamma f^{\prime}(E, 0),
\end{equation}
where $E$ is the particle energy.

Such a relation also holds for the general case of scattering of a relativistic particle by the scatterer (for example, by a nucleus) \cite{1}. 
According to \cite{1},
the forward scattering amplitude is related to the $T$-matrix describing the collision of particles in the general case as follows
\begin{equation}
\label{ins_18.10}
f(E, 0)=-(2\pi)^{2}\,\frac{m\gamma}{\hbar^{2}}\,T(E).
\end{equation}
From this we have
\begin{equation}
\label{insn_18.10}
f^{\prime}(E, 0)=-(2\pi)^{2}\,\frac{m}{\hbar^{2}}\,T(E).
\end{equation}
Using (\ref{18.10}), one can rewrite (\ref{18.9}) as follows
\begin{equation}
\label{18.12}
\vartheta=2\pi\rho\lambda_{\mathrm{C}}\texttt{Re}(f^{\prime}_{\uparrow\uparrow}-f^{\prime}_{\downarrow\uparrow})z,
\end{equation}
\begin{equation}
\label{18.13} \lambda_{\mathrm{C}}=\hbar/(mc)
\end{equation}
is the Compton  wavelength of the particle.

By means of the $T$-matrix, the expression for the rotation angle $\vartheta$ may also be represented in the form:
\begin{equation}
\label{ins_18.12}
\vartheta=-\frac{(2\pi)^{3}\rho}{\hbar c}\texttt{Re}(T_{\uparrow\uparrow}(E) - T_{\downarrow\uparrow}(E))z
\end{equation}
The path  length $z$ traveled by the particle in the target is $z = v t$, $v$ is the particle velocity, $t$ is the time in which the particle passed the path $z$. Hence, we also have $\vartheta=\omega_{\mathrm{pr}} t$, where $\omega_{\mathrm{pr}}$ is the particle spin precession frequency in a polarized target
\[
\omega_{\mathrm{pr}} =-\frac{(2\pi)^{3}\rho}{\hbar c}\, \ \texttt{Re}(T_{\uparrow\uparrow}(E)-   T_{\downarrow\uparrow}(E))v.
\]
As is seen, in a relativistic energy range the dependence $\vartheta\sim 1/k$ disappears, and the
entire possible dependence of the rotation angle $\vartheta$ on the particle energy is contained
in the amplitude $f^{\prime}(E, 0)(T(E))$.

According to \cite{nim06_PR+}, we have for a fully polarized
target:
\begin{equation}
\label{18.14} \vartheta\sim 10^{-3}\div 10^{-4}z,
\end{equation}
where $\vartheta$ is in radians, i.e., $\vartheta\sim 10^{-2}\div 10^{-3}$ rad
for a particle passing through a polarized target of length 10 cm.


With the increase in the target thickness, the influence of absorption on the polarization characteristics of a particle transmitted through matter is enhanced. Let, however, the target thicknesses be such that the spin-dependent contributions to the wave-function phase are small, i.e., inequalities $k \texttt{Re}g z \ll 1$ and $k \texttt{Im}g z \ll 1$ (see (\ref{ins1_18.5})) are fulfilled. In this case we have
\begin{equation}
\label{ins_18.14}
W^{\prime}=e^{ik\hat{n}z}W\simeq e^{ikn_{0}z}(1+i\frac{2\pi\rho}{k}(\vec{\sigma}\vec{g})z)W
\end{equation}
From (\ref{ins_18.14})
follows the expression for the number of particles $N$ transmitted through the target in the same direction as the direction of the momentum of the particles $N_{0}$ incident on the target  without being scattered.
\begin{equation}
\label{ins1_18.14}
N=N_{0}e^{-\rho\sigma z}[1-\frac{4\pi\rho}{k}\vec{P}_{0}\texttt{Im}\vec{g}z],
\end{equation}
where $\sigma$ is the spin-independent part of the total scattering cross-section determined by the imaginary part of $n_{0}$ in  (\ref{ins_18.5})
\begin{equation}
\label{ins12_18.14}
\sigma=\frac{4\pi}{k} \texttt{Im}(A+A_{3}\vec{n}\vec{n}_{1}+\ldots),
\end{equation}
$\vec{P}_{0}$ is the particle polarization vector before entering the target, $\vec{g}$ is defined by (\ref{ins1_18.5}).

According to (\ref{ins1_18.14}), the number of particles $N$
transmitted through the target  depends on the orientation of
$\vec{P}_{0}$: $N_{\uparrow\uparrow}\neq N_{\downarrow\uparrow}$,
where $N_{\uparrow\uparrow}$ describes $N$ for
$\vec{P}_{0}\uparrow \uparrow\texttt{Im} \vec{g}$,
$N_{\downarrow\uparrow}$ denotes $N$ for
$\vec{P}_{0}\downarrow\uparrow \texttt{Im} \vec{g}$.

So, spin dichroism occurs because the absorption coefficient of
incident particles in the target depends on the orientation of
their spin \cite{24a,PAX,PAX1}.

Using (\ref{ins_18.14}) for the spinor wave function $W^{\prime}$,
one can obtain the following expression for the polarization vector $\vec{P}$ of the particles transmitted through a polarized target:
\begin{equation}
\label{ins2_18.14} \vec{P}=\frac{\langle
W^{\prime}|\vec{\sigma}|W\rangle}{\langle
W^{\prime}|W\rangle}=\vec{P}_{0}+\frac{4\pi\rho z}{k}
\texttt{Im}((\vec{P}_{0}\vec{g})\vec{P}_{0}-\vec{g}) z
+\frac{4\pi\rho z}{k}[\vec{P}_{0}\times \texttt{Re}\vec{g}].
\end{equation}
In view of (\ref{ins2_18.14}), the polarization vector of high-energy particles undergoes rotation about the direction of $\texttt{Re}\vec{g}$. In a similar manner as in the case of low energies, the contributions associated with strong and P-, T-odd weak interactions can be distinguished by measuring the magnitudes of $N$ and $\vec{P}$ for different orientations of the polarization vector $\vec{P}_{0}$ of the particles incident on the target.

It is worthy of mention that for target nuclei with spin $J\geq 1$, in the cross-section $\sigma$, the term $\texttt{Im} A_{3}\vec{n}\vec{n}_{1}$  determined by the target quadrupolarization is different from zero.

In view of (\ref{ins1_18.14}),
the absorption of unpolarized ($\vec{P}_0=0$) antiprotons (protons) rotating in the ring (absorption of particle transmitted through the target)
will be different for different orientations of the quadrupolarization tensor of the deuteron target, i.e., the  lifetime of
a nonpolarized beam in the ring  depends on the orientation  of the quadrupolarization tensor of the target \cite{rouba}.
As a result, measuring  the beam lifetime in this case, one can determine
$\texttt{Im}A_{3}$, i.e.,  a spin-dependent part of the total cross-section of proton (antiproton)
scattering by a polarized  deuteron, which is proportional to $Q_{zz}$ .


\section{Proton (Antiproton) Spin Rotation  in a Thick Polarized Target}
\label{sec:protonspin}
With the increase in the target thickness the influence of the spin-dependent part of particle absorption in matter is enhanced.

In order to obtain equations describing  the evolution of intensity and polarization of a beam in the target, we shall split vector $\vec{g}$ into real and imaginary parts:
\begin{equation}
\label{ins_protonspin}
\vec{g}=\vec{g}_{1}+i\vec{g}_{2},
\end{equation}
where $\vec{g}_{1}= \texttt{Re} \vec{g}$; $\vec{g}_{2}= \texttt{Im} \vec{g}$.

Using (\ref{18.4}), (\ref{18.5}), one may obtain the following system of equations defining the relation between the number of particles $N(z)$ transmitted through the target  and their polarization $\vec{P}(z)$ \cite{nim06_PR+}:
\begin{equation}
\label{protonspin_1.51}
\frac{d\vec{P}(z)}{dz}=\frac{4\pi\rho}{k}[\vec{g}_{1}\times\vec{P}(z)]-\frac{4\pi\rho}{k}
\left\{\vec{g}_{2}-\vec{P}(z)[\vec{g}_{2}\vec{P}(z)]\right\}.
\end{equation}

\begin{equation}
\label{protonspin_1.52}
\frac{d N(z)}{dz}=-2\texttt{Im}(n_{0})k N(z)-\frac{4\pi\rho}{k}[\vec{g}_{2}\vec{P}(z)]N(z).
\end{equation}
(\ref{protonspin_1.51}) and (\ref{protonspin_1.52}) should be solved together under the initial conditions $\vec{P}(0)=\vec{P}_{0}$, $N(0)=N_{0}$. According to (\ref{protonspin_1.51}), the polarization of the incident particles passing through the polarized nuclear target undergoes rotation through the angle
\begin{equation}
\label{protonspin_1.53}
\theta=\frac{4\pi\rho}{k}g_{1} z.
\end{equation}

Assume that the target polarization $\vec{P}_{t}=\frac{\langle\vec{J}\rangle}{J}$ is parallel (this is the case of longitudinal polarization)
or orthogonal (transversal polarization) to the momentum of the incident particle $\vec{k}$.
Then $\vec{g}_{1}\parallel\vec{g}_{2}\parallel\vec{P}_{t}$ and  (\ref{protonspin_1.51}), (\ref{protonspin_1.52}) are
reduced to a simple form.

Consider two specific cases for which the initial polarization of
the incident beam is (a) $\vec{P}_{0}\parallel \vec{P}_{t}$ or (b)
$\vec{P}_{0}\perp\vec{P}_{t}$.

Case (a) is a standard transmission experiment  wherein we observe
the process of absorption in the polarized target without a change
in the direction of the initial beam polarization. The absorption
is different for particles polarized parallel and antiparallel to
the target polarization. The number of particles $N$  changes
according to
\begin{equation}
\label{protonspin_1.55}
N(z)=N_{0}\exp(-\sigma_{\pm}\rho z),
\end{equation}
where
\begin{equation}
\label{protonspin_1.56}
\sigma_{\pm}=\frac{4\pi}{k}\left[\texttt{Im}(A+A_{3}\vec{n}\vec{n}_{1})\pm\texttt{Im}A_{1}J P_{t}\pm \texttt{Im} A_{2} J P_{t}\right].
\end{equation}
In case (b) the coherent scattering by the polarized nuclei results in spin rotation of the incident particles about the target polarization $\vec{P}_{t}$.

According to (\ref{18.9}), the spin rotation angle
\begin{equation}
\label{ins.protonspin_1.56}
\vartheta=\frac{4\pi\rho}{k} \texttt{Re} g=\frac{4\pi\rho}{k}\left[\texttt{Re}A_{1} J P_{t}+\texttt{Re}A_{2} J (\vec{n}\vec{P}_{t})\right]z
\end{equation}
is directly connected with the real part of the forward scattering amplitudes.

The values of $\texttt{Re}A_{1}$  and $\texttt{Re}A_{2}$ can be determined separately by measuring spin rotation angles for two cases when the target spin is parallel and antiparallel to the beam direction $\vec{n}$. This means that by measuring the final intensity and polarization of the beam in cases (a) and (b)  we can directly reconstruct the spin-dependent forward scattering amplitude.

Let us assume now that the the target polarization is directed at some angle
(which does not equal $\pi/2$ ) with respect to the incident particle momentum
 and that the incident beam polarization is perpendicular to the plane formed
 by the vectors $\vec{P}_{t}$ and $\vec{n}$ (see Fig. 1). In this case the effect of proton (antiproton)
 spin rotation about the vector $\vec{g}_{1}$ combined with absorption dichroism,
 determined by the vector $\vec{g}_{2}$, will cause the dependence of the total number
 of particles transmitted through the target on $\texttt{Re} A_{1}$ and  $\texttt{Re} A_{2}$ \cite{nim06_PR+,21a}:

\begin{figure}[h]
\epsfxsize = 6 cm \centerline{\epsfbox{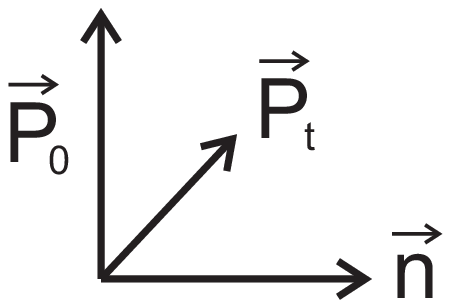}}
\caption{Figure 1} \label{fig1}
\end{figure}

\begin{equation}
\label{ins.protonspin_1.56+}
N(z)\sim(\texttt{Re}A_{1}\cdot\texttt{Im}A_{2}-\texttt{Re}A_{2}\cdot\texttt{Im}A_{1})\left[\vec{P}_{t}
\times\vec{n}(\vec{n}\vec{P}_{t})\right]\vec{P}_{0}
\end{equation}
Such behavior of $N(z)$ enables measuring spin-dependent contributions to the amplitude $f(0)$
in the transmission experiment without measuring the polarization of the beam transmitted through the target \cite{nim06_PR+,21a}.

This implies that if a vertically polarized beam rotates in the storage ring (the beam's spin is
orthogonal to the orbit plane), then the decrease in the beam intensity (the beam lifetime in the storage ring)
in the case when vectors $\vec{P}_{t}$ and $\vec{n}$ lie in the horizontal plane (in the orbit plane) enables
determining $\texttt{Re} A_{1}$ and  $\texttt{Re} A_{2}$.

Note that the contribution under study reverses sign when vector
$\vec{P}_{t}$  rotates about $\vec{n}$ through $\pi$ (or
$\vec{P}_0$ changes direction: $\vec{P}_0\rightarrow -\vec{P}_0$).
Measuring the difference between the beam's damping times for
these two orientations of $\vec{P}_{0}$ ($\vec{P}_{t}$),
 one can find the contribution of $\texttt{Re} A_{1}$ and  $\texttt{Re} A_{2}$ to the real part of the amplitude $f(0)$.

Unlike a spin rotation experiment, this transmission experiment does not allow us to determine
$\texttt{Re}A_{1}$ and $\texttt{Re}A_{2}$ separately.

There is also an inverse process: if a nonpolarized beam of particles is incident on
the target, then after passing through the target, the beam acquires polarization orthogonal to the plane formed by $\vec{P}_{t}$
and $\vec{n}$. This phenomenon is most interesting in the case of a storage ring. If $\vec{P}_{t}$ and $\vec{n}$ lie in the orbit
plane of a circulating beam (i.e., the horizontal plane), then in the course of time the beam acquires vertical polarization
orthogonal to the orbit plane. By measuring the arising vertical polarization it is also possible to determine the spin-dependent
part of the  coherent elastic zero-angle scattering amplitude. In particular, in spin-filtering experiments on obtaining polarized
beams of antiprotons (protons), it is sufficient to perform an experiment under the conditions when the polarization vector $\vec{P}_t$
of a gas target is directed at a certain angle, which is not equal to $0,\, \pi$ or $\pi/2$, with respect to the particle momentum
direction $\vec{n}$. If $\texttt{Re}A_{1,2}\sim \texttt{Im}A_{1,2}$,
the degree of arising polarization is comparable to the anticipated degree of polarization of the anti(proton) beam arising in the
PAX method and enables one to measure the contribution proportional to $\texttt{Re}A_{1,2}$.

Worthy of mention is that in the case under consideration the beam
lifetime depends on the orientation of $\vec{P}_t$ in the
$\vec{P}_t,\vec{n}$-plane. In this case the measurement of the
beam lifetime also gives information about $\texttt{Re}A_{1,2}$.


\section{Influence of multiple Coulomb
scattering on "optical"
spin precession and dichroism of charged particles in
pseudomagnetic  fields of polarized target} \label{protonspin_p}

It should be emphasized, however, that the foregoing
simple picture of the spin rotation phenomenon  holds only for
neutrons. In the case of charged particles (protons, antiprotons,
nuclei) the collision process is determined not only by
short-range nuclear interactions but also by a long-range
electromagnetic interaction of  incident particles with the
Coulomb field of the target particles.  The magnitude of the total
scattering cross-section $\sigma_{N}$ associated with nuclear
interaction is of the order of $\sim 10^{-24}$ cm. Hence, it
follows that the mean free path
\[
l\sim\frac{1}{N\sigma_{N}}\sim 10^{2}\, \textrm{cm}
\]
when the particle number density $N$ in the target is $\sim
10^{22}$ cm$^{-3}$.

At the same time, the cross-section for Coulomb scattering by a
screened potential produced by the target nucleus is
\[
\sigma_{\mathrm{C}}\simeq \frac{2 \pi m
R^{2}e^{4}Z_{N}^{2}}{\hbar^{2} E},
\]
$E$ is the particle energy, $R$ is the shielding radius, $Z_N|e|$ is
the charge of the nucleus, $e=\pm |e|$ is the charge of the
particle, $|e|$ is the magnitude of the electron charge. As a
result we have for particles with energies of tens-hundreds of
MeV:
\[
l_{\mathrm{C}}\simeq \frac{1}{N\sigma_{\mathrm{C}}}\simeq
\frac{1}{Z_{N}^{2}}10^{-4}\div 10^{-3}\, \textrm{cm}\ll l.
\]

As is seen, over the nuclear mean free path $l$,  a charged
particle undergoes a large number of collisions.

It seems that this effect should lead to practically total
suppression of coherent spin rotation of charged particles in the
polarized nuclear target, but this is not the case. It turns out
that a scattered particle also "senses" the quasimagnetic
(quasielectric) nuclear field in which its spin is rotated.
Coulomb scattering in the nuclear target results in only
insignificant depolarization of the beam and does not influence
coherent spin rotation \cite{nim06_PR+}. Moreover, spin rotation
for charged particles changes since, besides coherent spin
rotation in the quasimagnetic  (quasielectric) nuclear field,
there is incoherent spin rotation arising in  scattering events by
the target nuclei. This incoherent spin rotation is conditioned by
the interference between the nuclear and the electromagnetic
interaction  \cite{nim06_PR+,B9,B10} and incoherent
particle-nucleus scattering \cite{B9}.

Study of the rotation angle and spin dichroism of the beam in the
interval of small angles with respect to the incident direction of
the initial beam enables one to explore the contributions to the
effect, which are  associated with coherent and incoherent
scattering \cite{bar+shirv}. Interestingly enough, in the case
when the rotation effect is measured by the detector recording
particles scattered at all angles ($4\pi$-geometry), in the
high-energy range in the first-order Born approximation to the
Coulomb interaction, the contribution to the spin rotation angle
of the beam, which is due to the contribution of the Coulomb
interaction to the spin-dependent part of the zero-angle
scattering amplitude, and the contribution to the spin rotation
angle, which is due to the interference between nuclear and
Coulomb interactions, compensate each other \cite{B10,bar+shirv}.
Using spin density matrix formalism, one can carry out a  thorough
study of  spin rotation angle and  spin dichroism effect under the
conditions when multiple Coulomb and nuclear scattering are
essential. Next, we shall briefly recall the procedure for the
derivation of the equation for spin density matrix
\cite{nim06_PR+,B9,B10,bar+shirv}.

\textbf{Spin Density Matrix Formalism}

Under the conditions when multiple scattering appears important,
quantum mechanical description of a beam of polarized particles
traveling through a polarized target utilizes the spin density
matrix $w$.


In our case, we treat the polarized target as a thermal reservoir
with an infinite set of degrees of freedom which we then average.
Therefore the density matrix for the system, $\hat{\rho}(b, T,
t)$, consisting of the polarized target plus polarized particle
beam, can be written as a tensor product of the matrices
$\hat{\rho}_b (t)$  and $\hat{\rho}_T(t)$ ($\hat{\rho}_b(t)$
is the beam spin density matrix, $\hat{\rho}_T(t)$ is the target
spin density matrix):
\begin{equation}
\label{protonspin_1.12} \hat{\rho} (b, T, t)=\hat{\rho}_b(t)\otimes \hat{\rho}_T (t).
\end{equation}
The master equation for the spin density matrix of this system has
a form \cite{23a}
\begin{equation}
\label{protonspin_1.13}
\frac{d\hat{\rho}_b(t)}{dt}=-\frac{i}{\hbar}(\hat{H}\hat{\rho}_b-\hat{\rho}_b\hat{H})+
\left(\frac{d\hat{\rho}}{dt}\right)_{\mathrm{sct}}
\end{equation}
with the Hamiltonian $\hat{H}$ including the interaction between
particles of the beam and the external electromagnetic fields. The
term $(d\hat{\rho}/dt)_{\mathrm{sct}}$  describes the change in
the density matrix due to collisions in the target.

A detailed account of general methods for obtaining the explicit
form of the collision integral is given in many manuals
\cite{23a}. They can also be applied to the consideration of the
interaction between  polarized beams and a polarized target
\cite{nim06_PR+}.  We shall first make use of the fact that in the
case of high-energy particles, the energies of the particles are
much greater than the binding energies of the scatterers with the
target.   This enables employing impulse approximation \cite{1},
so we may consider the collisions between the beam of particles
and the target scatterers resting at points $\vec{R}_{n}$  of
their instantaneous coordinates of centers of mass
\cite{nim06_PR+,bar+shirv}.

If the particle energy loss through collision with the target nuclei can be neglected, the equation for the density matrix describing the
behavior of the polarized particle beam in polarized nuclear
matter can finally be written as \cite{nim06_PR+}:
\begin{eqnarray}
\label{protonspin_1.22}
\frac{d \hat{\rho}(k)}{d t}=& &-\frac{i}{\hbar}[\hat{H},\rho]+v N \,Tr_{\mathrm{T}}\left[\frac{2\pi i}{k}\left[\hat{F}(\theta=0)\rho(\vec{k}) -\rho(\vec{k})\hat{F}^{+}(\theta=0)\right]\right.\nonumber\\
& &\left.+ \texttt{Tr}_T\int d\Omega_{\vec{k}^{\prime}}
F(\vec{k}, \vec{k}^{\prime})\rho(\vec{k}^{\prime})F^{+}(\vec{k}^{\prime},\vec{k})\right],
\end{eqnarray}
where $|\vec{k}^{\prime}|=|\vec{k}|$, $v$  is the speed of
incident  particles, and $N$ is the density of nuclei in the
target. In (\ref{protonspin_1.22}) we take the trace over the spin
states of the target nuclei, $\hat{F}$ is the scattering
amplitude.

It should be noted that, strictly speaking,
 otherwise, in analyzing multiple scattering of particles in a target containing
 light nuclei (protons, deuterons), one should take account of the energy loss of
 the incident particle through a single collision with the target nucleus
 \cite{bar+shirv}. The density matrix equation for this case see
 in \cite{bar+shirv}.

Equation (\ref{protonspin_1.22}) was analyzed in
\cite{nim06_PR+,B10}, where  two physical mechanism of spin
rotation were indicated: one due to refraction of particles in a
polarized target ("optical" spin rotation) (the second term in
(\ref{protonspin_1.22}))
 and the other, appearing as a result of spin rotation through incoherent scattering and caused by Coulomb-nuclear
 interaction and incoherent scattering by nuclei \cite{nim06_PR+,B9,B10,bar+shirv}
 (the third term in (\ref{protonspin_1.22})) [See \cite{bar+shirv}].  As is shown in \cite{bar+shirv}, using different angular
 resolution of the detector,  one can  study different contributions  to spin rotation.

Thus, multiple scattering does not cancel the effect of proton
(antiproton) spin rotation and spin dichroism in a polarized
target.

\section{Conclusion}

It was shown that in current spin-filtering experiments to produce
polarized antiprotons (protons, deuterons), it is possible to
measure a real part of a coherent elastic  amplitude of proton
(antiproton, deuteron) scattering at zero angle, when the plane
formed by the target polarization vector $\vec{P}_t$ and the beam
momentum direction $\vec{n}$ lies in the orbit pane of the beam
and the angle between $\vec{P}_t$ and $\vec{n}$ differs from 0,
$\pi$, or $\pi/2$. Anticipated degree of polarization in this
geometry can be compared to that of antiprotons (protons)
(anticipated change in the beam lifetime with the change in the
orientation of $\vec{P}_t$ in the ($\vec{P}_t,\vec{n}$) plane),
which is observed with the spin-filtering method  \cite{PAX,PAX1}.

\end{document}